# The underlying dynamics of life and its evolution: A Prigogine-Inspired Informational Dissipative System

Salvatore Chirumbolo[1]*, Antonio Vella[2]

[1]Department of Engineering for Innovation Medicine, University of Verona, Italy

[2]Unit of Immunology, AOUI Verona, Italy

Correspondence:

Prof. Salvatore Chirumbolo, PhD

Department of Engineering for Innovation Medicine

University of Verona

Strada Le Grazie 8, 37134 Verona (Italy)

Tel +390458027645

e-mail salvatore.chirumbolo@univr.it

The authors state they have no conflict of interest

## Abstract

Life, in its foundation, is a puzzling conundrum for science. Understanding the interplay between chaos, entropy dynamics, and Prigogine's dissipative systems provides profound insights into the emergence, stabilization, and eventual collapse of far-from-equilibrium systems. It is plausible that, intertwined with the thermodynamic dissipative systems, highlighted by Ilya Prigogine, informational dissipative systems as well actively works to provide inanimate matter with the typical properties of living objects, such as autopoiesis. This study explores the cyclic entropy flows between water topology (Shannon space) and molecular systems (Boltzmann space), focusing on the critical role of disquisotropic entropy—an informational entropy reservoir formed by imperfections and variability in molecular systems. Our analysis reveals that chaos, far from destabilizing the system, acts as a stabilizing force, enhancing resilience, adaptability, and longevity by delaying thermodynamic equilibrium.

This work bridges foundational thermodynamic principles with the emergent behaviour of chaotic systems, opening pathways to a deeper understanding of complexity in nature and technology. By elucidating the interplay of chaos, entropy, and dissipative dynamics, we advance a paradigm where disorder becomes a tool to sustain order, a hallmark of life and the essence of complex systems.

## Introduction

A recent paper by Diego Gonçalves in the journal *Theory in Biosciences*, addressed

the puzzling conundrum of the foundations of life (Gonçalves, 2024).
The author proposes a fresh perspective by describing life as a combination of two fundamental dissipative mechanisms: the creation of chemical disequilibria and their self-constrained dissipation. This universalizes life as a thermodynamic phenomenon, breaking away from traditional biology-centric definitions (Gonçalves, 2024). The use of the second law of thermodynamics to frame life processes provides a robust physical basis, bridging biology and physics. It elegantly ties concepts like entropy production and free energy conversion to life origin. We also observed that the article outlines scenarios for recreating life experimentally, emphasizing electron bifurcation as a key mechanism. This testability strengthens the scientific value of the paper. Moreover, by involving concepts from biochemistry, physics, and evolutionary biology, the paper has broad interdisciplinary relevance.

In addition, it speaks to audiences in origin-of-life research, astrobiology, and theoretical biology. The critique of Darwinian evolution as a criterion for defining life is thought-provoking. It offers an expanded understanding of selection that transcends self-replication. Despite all these issues, we highlighted some weakness in the paper.

While the theory is compelling, it relies heavily on theoretical constructs without presenting experimental data or models validating its hypotheses. By reducing life to dissipative processes, the paper risks overlooking the complexity of cellular systems, including genetic and epigenetic regulation, and their interactions with environments. The focus on hydrothermal vents and electron bifurcation discounts other plausible origins of life scenarios, such as ice matrix catalysis or extra-terrestrial panspermia. Terms like "self-constrained dissipation" and "entropic dams" are introduced without sufficient operational definitions or mathematical formalism, which could alienate readers outside thermodynamics. We found that critics of thermodynamic models for life, such as those emphasizing the irreducible complexity of biological systems, are not addressed. Major novelties we found regarded the following issues. The paper introduces a thermodynamic definition of life, where aliveness is quantified by the degree of self-constrained dissipation. Identifying electron bifurcation as the cornerstone for creating chemical disequilibria at life origin offers a novel focal point for experimental studies. The argument that self-replication is a byproduct rather than a defining feature of life challenges traditional definitions and opens new avenues for synthetic biology. The call for engineering artificial electron bifurcation systems provides clear steps for experimentally testing the theory (Gonçalves, 2024).

The effort published by the author (Gonçalves, 2024) strengthens the idea that understanding how biological life is organized at its deepest foundations represents a scientific challenge that has spanned millennia. At present, many researchers, seeking to uncover the intricate details that might enable us to establish objective and indisputable criteria for distinguishing between inanimate and living structures, have proposed various hypotheses and theories (Hazen, 2017; Spitzer et al., 2015; Miller et al., 2023; Chirumbolo and Vella, 2023; Mikhailovsky, 2024; Heylighen et al., 2022). Fundamentally, however, they all follow a similar path.

Briefly speaking, chemical molecules adhere to the laws of "necessity" and "chance," which, on their own, are thought to constitute the driving force behind the dynamics that characterize a biological entity and set it apart from a non-biological one (Carroll, 2001; Solé et al., 2024). However, chance and necessity are antithetical and opposing realities, one possessing the power to disrupt and destroy what the other constructs. Again, the latter would have no freedom of action except within a hypothetical and highly improbable teleonomy. Biological systems are ordered systems, despite their apparent disorder, which we often describe using terms such as chaos or complexity. Notwithstanding, the cosmic underpinning of these concepts is far more nuanced and richer than it might initially seem. A new language and a new model are essential to interpreting biological phenomena in their deepest essence.

On the other hand, Prigogine's dissipative structures are a cornerstone in understanding self-organization in systems far from equilibrium, and their application in the studies of Kondepudi and Dixon explores real-world manifestations of these ideas in biology, besides chemistry (Kondepudi et al., 2020; Kondepudi et al., 2017a; de Bari et al., 2024).

Yet, the theoretical foundation of necessity and chance in the context of dissipative structures lies at the intersection of non-equilibrium thermodynamics, nonlinear dynamics, and probability theory. Prigogine's framework provides insights into how order emerges and evolves in systems far from equilibrium, influenced by deterministic laws (necessity) and stochastic fluctuations (chance).

Necessity refers to the deterministic processes and constraints imposed by physical laws, driving the system toward certain outcomes (Solé et al., 2024). In a "necessity" scenario, we could highlight key concepts such as: a) the system is governed by nonlinear differential equations, such as reaction-diffusion equations or the Navier-Stokes equations; b) systems evolve to states that maximize entropy production within constraints, according to the Prigogine theorem (Kondepudi et al., 2017b); c) deterministic feedback mechanisms lead to stable attractors or transitions (bifurcations) to new ordered states. In this context, it appears that necessity ensures that a) dissipative structures require energy gradients to maintain their order, b) that far-from-equilibrium conditions lead to deterministic patterns of self-organization, such as convection rolls or chemical oscillations, and that c) under identical conditions, the system follows deterministic trajectories (e.g., fixed points, limit cycles). An example is the Rayleigh-Bénard convention, inasmuch when a fluid is heated from below, deterministic laws (e.g., fluid dynamics, thermal gradients) drive the formation of convection cells.

This vision of necessity enabled to elucidate Prigogine's dissipative structures, seems to be suitable also for chance definition.

Chance refers to stochastic fluctuations or random perturbations that influence system behaviour, often arising from molecular-scale randomness or environmental noise. Therefore, chance can be defined by key concepts such as: a) random fluctuations at the microscopic level (e.g., Brownian motion); b) noise can amplify or stabilize patterns in systems near critical thresholds; c) chance can push a system to one of several possible states, even when deterministic laws allow multiple outcomes. Therefore, chance introduces fluctuation-driven transitions, i.e., that random perturbations can trigger transitions between states, particularly near bifurcation points. Moreover, stochastic effects lead to different outcomes in otherwise identical systems (e.g., symmetry breaking in chemical chirality). And finally, in biological systems, chance enables evolutionary novelty and adaptability. An example can be found in the chiral symmetry breaking, inasmuch in sodium chlorate crystallization thermal noise determines whether left- or right-handed crystals dominate.

Prigogine argued that the interplay of necessity and chance offers a new perspective on nature. Far-from-equilibrium conditions reveal the limits of classical determinism. The emergence of order from chaos highlights the irreversibility of time, blending deterministic and stochastic processes. Chance allows nature to "choose" between multiple possibilities, fostering diversity and innovation.

However, Prigogine introduced the concept of dissipative structures to explain how systems maintain order through continuous energy exchange with their environment; in other words, how “disorder” is able to create “order”. These structures are characterized by self-organization and arise in non-equilibrium conditions, leading to the emergence of complex patterns and behaviours. It is assumed that Prigogine's theories provide a framework for understanding how complex systems, such as biological organisms and ecosystems, self-organize and evolve over time (Kondepudi et al., 2017b; Goldbeter, 2018).

In this tale, the fundamental idea that dissipative structures, in biology, should obey to a mandatory dynamic ruled by information, is quite dismissed from any sound scientific debate, as organisms are biased in their simplest interpretation as complex automata or machineries.

Dissipative structures are ordered states that emerge in non-equilibrium systems through energy dissipation, yet this dissipation is mainly of informational nature. They require continuous energy input to sustain their order but they are not merely thermodynamic machines.

Both Kondepudi and Dixon emphasized the role of nonlinear feedback in the formation and stability of dissipative structures. They studied how systems undergo bifurcations, leading to new states of order or chaos under changing external conditions (Kondepudi et al., 2020; Kondepudi et al., 2017a; De Bari et al., 2024). Their contribution to the deepest comprehension of life emergence and evolution is paramount. Prigogine’s framework, as extended by Kondepudi, helps explain how life might have originated as a dissipative structure driven by energy flows on prebiotic Earth.

Anyway, so far, dissipative structures have been discussed in the chemical-physical and thermodynamic contexts, while it is not impossible to hypothesize that dissipative systems, as described by Prigogine, could instead be fundamentally informational dissipative systems. This could be due to the high informational availability of water (in a topological sense) and the relationship between information and energy according to Landauer's principles. In this work, we will delve deeper into these aspects to explain how life originated and autonomously developed. First, we will attempt to answer the question of why almost all complex systems derived from dissipative structures are related to chaos or which we trivially define "disorder" (Chirumbolo and Vella, 2023, Chirumbolo and Vella, 2021).

### Prigogine's dissipative structures becoming chaotic. Role of water

Ilya Prigogine's dissipative structures describe systems far from thermodynamic equilibrium that self-organize by dissipating energy. Chaos emerges in such systems when certain parameters, such as energy input or interaction rates, cross critical thresholds, leading to unpredictable behaviour.

To explore this transition mathematically and graphically, let us use the Lorenz system as a representative of dissipative structures. The Lorenz system models the dynamics of convection rolls in fluid dynamics but can represent any dissipative system transitioning to chaos. Therefore, we can define the equations, as the Lorenz system has three coupled nonlinear differential equations:

(1) $$\frac{dx}{dt} = \sigma(y - x)$$

(2) $$\frac{dy}{dt} = x(\rho - z) - y$$

(3) $$\frac{dz}{dt} = xy - \beta z$$

where $x$, $y$ and $z$ are state variables (e.g., representing system components), ρ is the Prandtl number (which controls energy dissipation rate) (Yigit et al., 2020), ρ is the Rayleigh number (energy input to the system) and β the aspect ratio (geometry-dependent dissipation). Transition to chaos occurs following these considerations: a) at low ρ, the system exhibits stable fixed points (no chaos), whereas as ρ increases, bifurcation occurs, leading to chaotic behaviour for certain parameters ranges. After solving the Lorenz equations numerically, we can plot the trajectory in 3D to visualize chaos and plot time series to show unpredictability in the chaotic regime, using a Python code with Matlab (Figure 1).

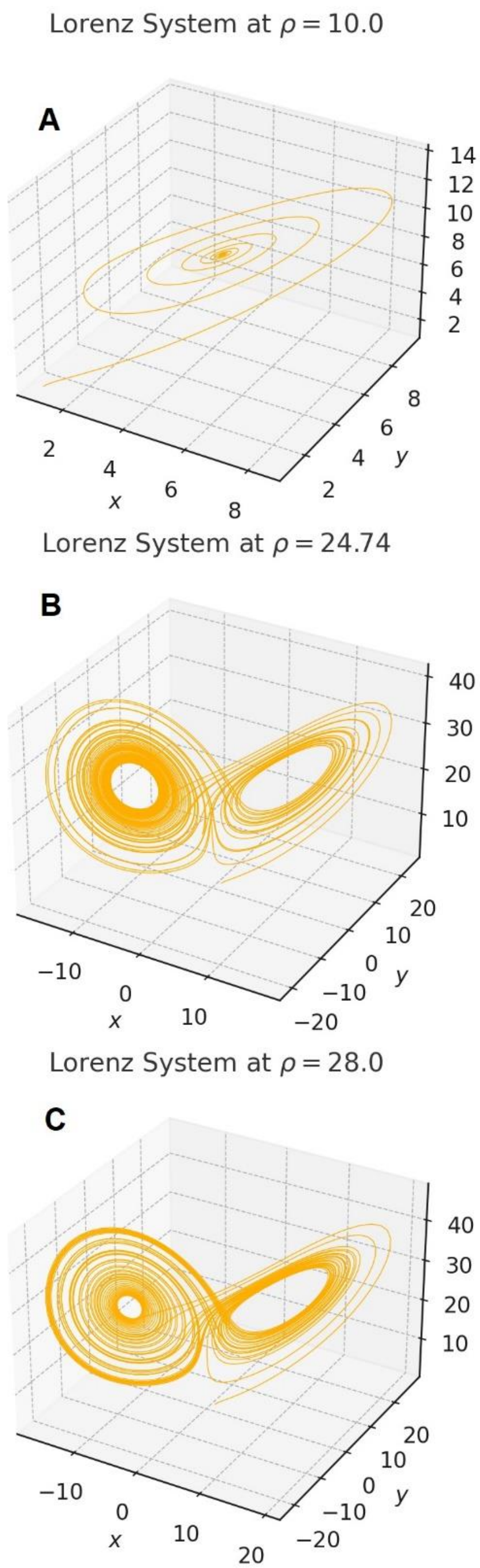
Lorenz System at $\rho = 10.0$
A
x
y
Lorenz System at $\rho = 24.74$
B
x
y
Lorenz System at $\rho = 28.0$
C
x
y

Figure 1

In biology, these systems need to be thoroughly revised by the involvement of two major components, of utmost importance: water and information.

Introducing the topological role of water molecules and their contribution to dynamic information flow adds an additional layer of complexity to the system (Chirumbolo and Vella, 2023). Water molecules in biological and chemical systems act as a topological field. Water forms hydrogen bond networks, which can dynamically adapt and reconfigure based on external inputs (e.g., energy, solutes, or fields), so allowing liquid water to behave as a topological field, able to provide molecules with a huge deal of probabilities to change their intrinsic informational endowment, once they are in water.

Furthermore, these dynamic rearrangements influence energy dissipation pathways and even the effective "communication" between molecules (Ball, 2008). Water molecules store, release, and transmit energy through vibrational, rotational, and translational modes. This modulation can affect how energy flows into the system (represented by in the Lorenz equations). The network of water molecules can change the effective dimensions or constraints of the system, leading to emergent behaviours (like periodicity or chaos).

Now, by introducing a time-varying, we should have that (energy input) becomes a dynamic parameter instead of a constant, fluctuating due to water reorganizations. Therefore, let ρ(t) be a function of time, reflecting dynamic water-mediated information flow. Furthermore, water topology depends on the system state variables $(x,y,z)$.

So:

(4) $$\rho(t) = \rho_0 + \Delta\rho\,(\sin \omega t + \phi)\, f\,(x,y,z)$$

where:

$\rho_o$ = baseline energy input;

$\Delta\rho$ = amplitude of dynamic variation;

$\omega$ = frequency of water topology oscillations;

$f\,(x,y,z)$ = state-dependent feedback

Dynamic ρ$(t)$\ introduces non-stationarity to the Lorenz system, potentially increasing complexity and driving new transitions between order and chaos.

We modify the Lorenz equations by making dynamic:

(5) $$\frac{dx}{dt} = \sigma(y - x)$$

(6) $$\frac{dy}{dt} = x\,(\rho(t) - z) - y$$

(7) $$\frac{dz}{dt} = xy - \beta z$$

(8) $$\rho(t) = \rho_0 + \Delta\rho\,(\sin \omega t + \phi)(1+\frac{z}{z+1})$$

Figures 2 and 3 represent the plotting of these dynamics

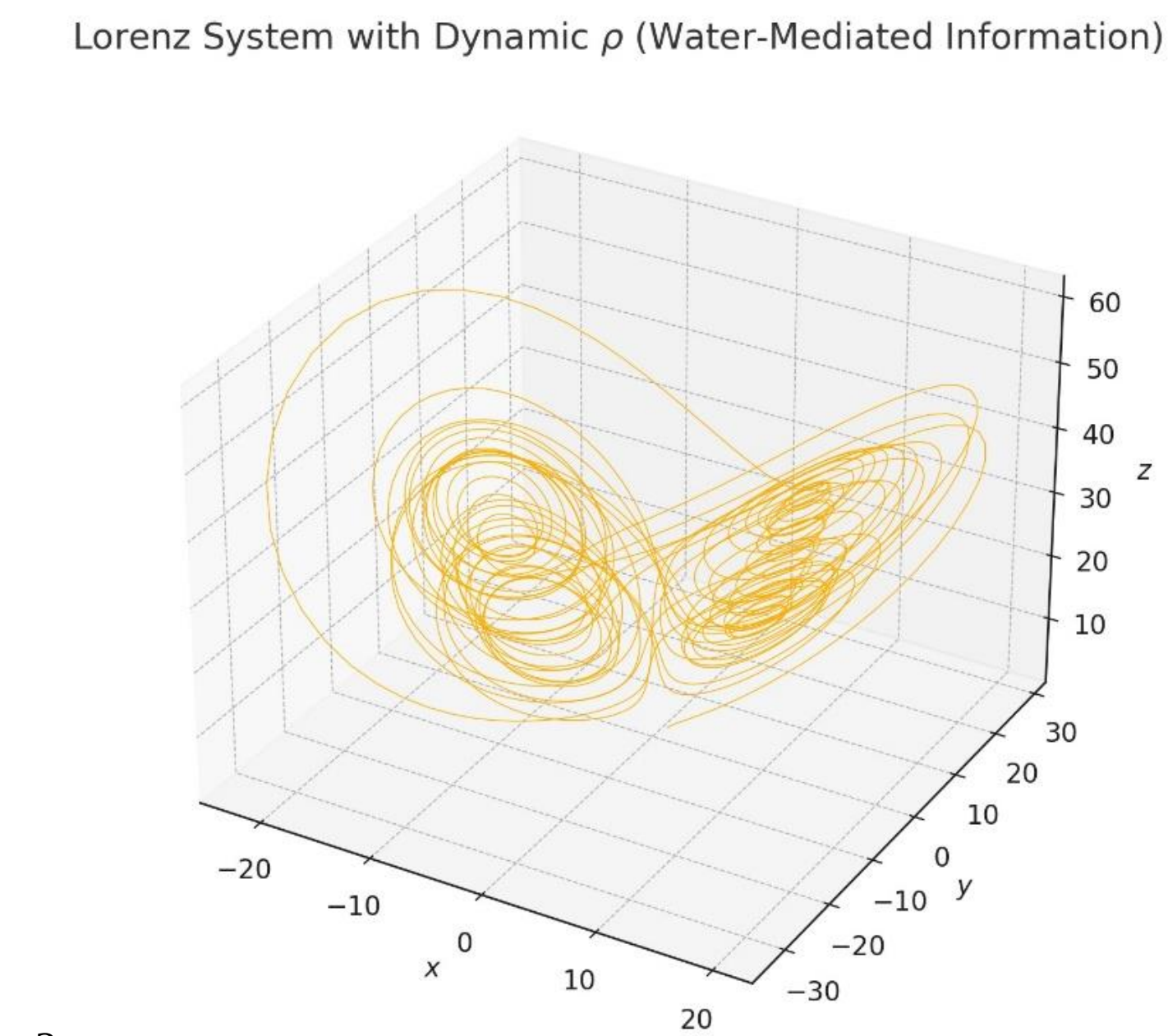

Figure 2

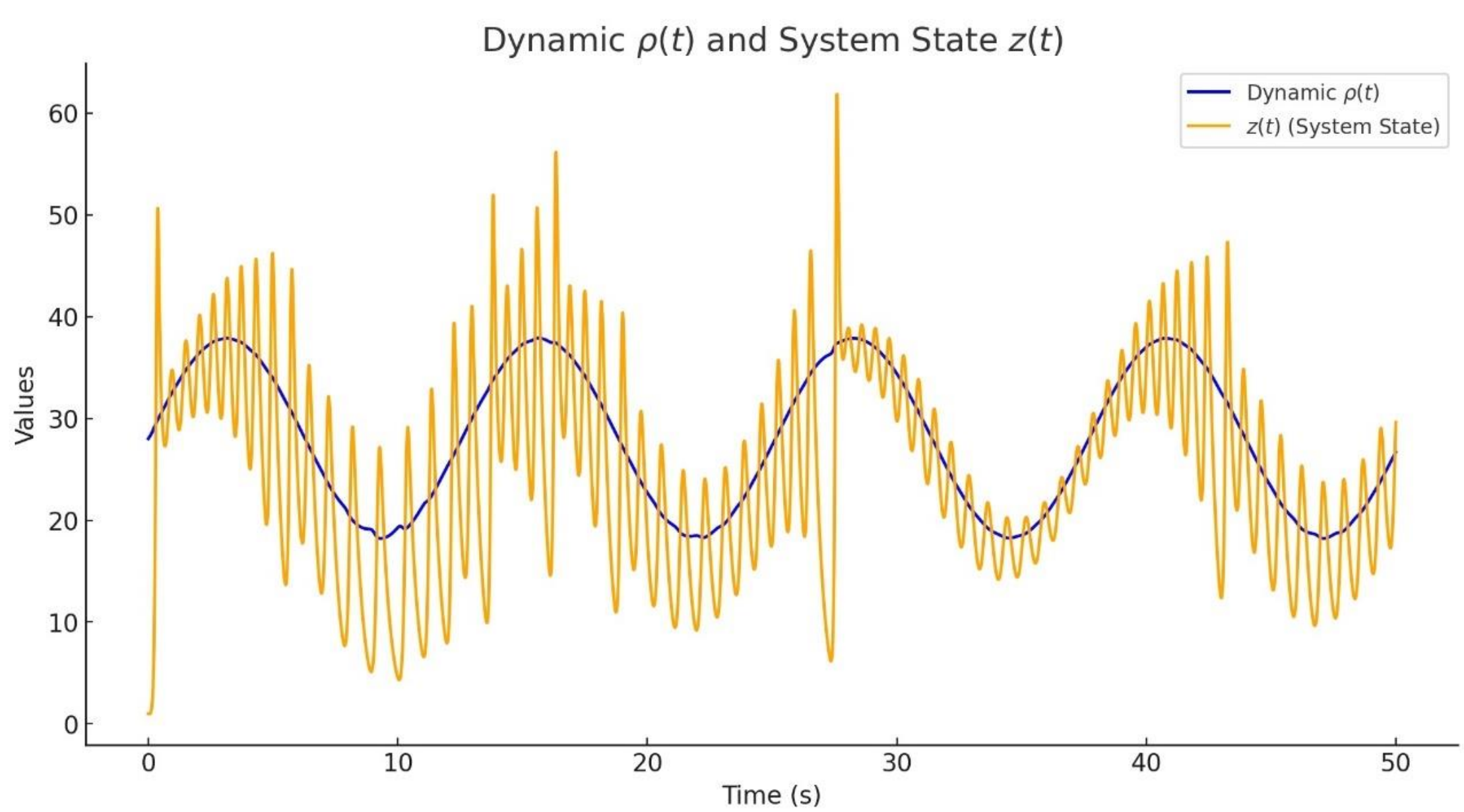

Figure 3

This modified scenario suggests that water-mediated dynamic information flow significantly amplifies complexity, introducing unpredictable behaviour and coupling between energy inputs and system states.

The topology of water, i.e., the network of hydrogen bonds, molecular interactions, and structural configurations of water molecules, affects chaotic systems in profound ways.

Let us tackle both numerical simulation and further exploration of how water topology affects chaotic systems. We should modify the Lorenz system to introduce stochastic noise and time-dependent parameters that mimic water dynamic topology. And moreover, we should include random fluctuations to represent quantum noise or thermal vibrations in the water network. Again, we have to visualize the altered trajectories, time series, and phase space. In this context, we should discuss broader implications, such as biological and physical systems where water topology critically affects chaotic behaviour.

Figure 4 and 5 are possible plots of these concepts.

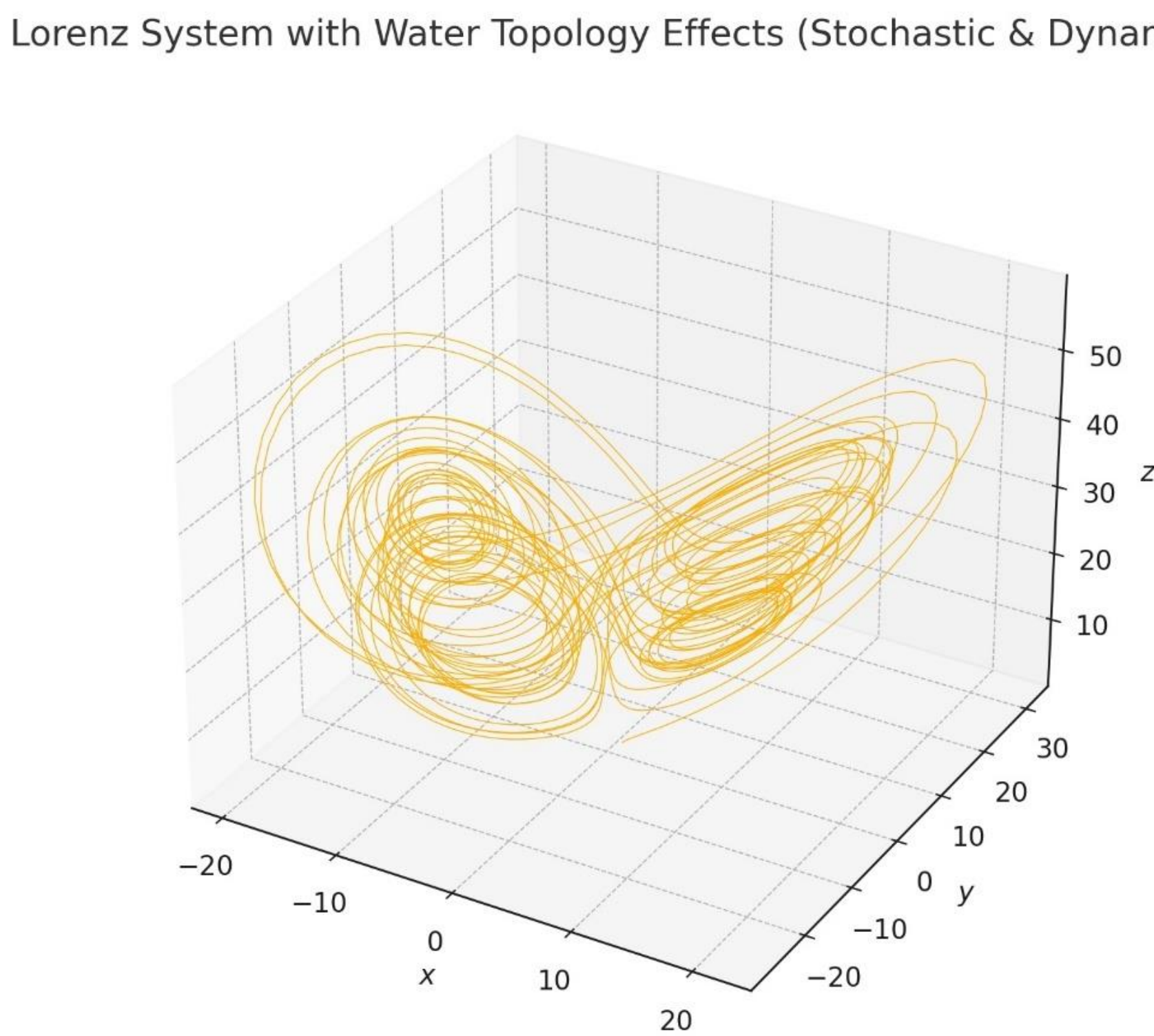

Figure 4

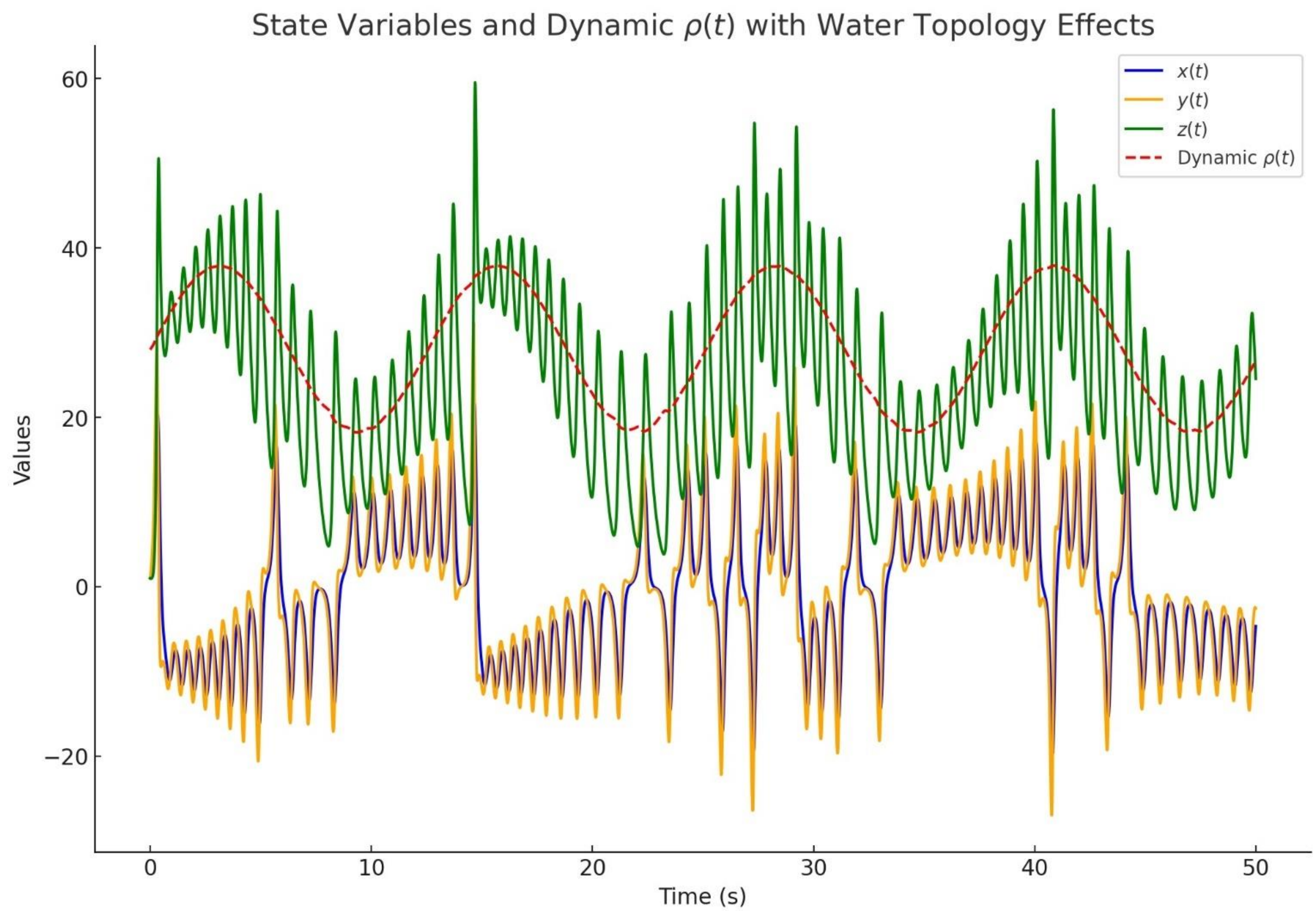

Figure 5

Regarding the relationships between water and the molecular states of any molecule in water, the molecular states follow a system of nonlinear equations (similar to a Lorenz-like system) with modulation by the water topology state *W(t)*:

(9) $$\frac{dx}{dt} = \sigma(y - x) + \alpha W$$

(10) $$\frac{dy}{dt} = x(\rho(t) - z) - y$$

(11) $$\frac{dz}{dt} = xy - \beta z$$

where:

$x(t)$ represents molecular state dynamics such as molecular displacements or reconfigurations;

$y(t)$ reflects energy exchange involving forces and interactions between molecules,

$z(t)$ captures entropy dissipation, showing how molecular systems lose or redistribute energy over time. The interplay with $W(t)$ is shown by the parameter of coupling ($\alpha W$), which indicates that the water topology modulates $x(t)$ amplifying or dumping its oscillations and by the parameter of dynamic $\rho(t)$, which indicates that the energy input $\rho(t)$ is dynamically influenced by $W(t)$, creating nonlinear feedback that propagates through $x(t)$, $y(t)$ and $z(t)$.

As about the water topology state, it evolves according to its nonlinear dynamics, interacting with molecular states:

$$\frac{dW}{dt} = \gamma(S - W) - \delta W^2 \tag{12}$$

where:

$\gamma(S\text{-}W)$, reflects coupling between molecular states (S) and water topology (W);

$-\delta W^2$, captures the non-linear damping of water topology representing structural reorganizations. Actually, peaks in $W(t)$ occur when molecular dynamics require stabilization or adjustment. These peaks align with transitions or critical changes in $x(t)$, $y(t)$, $z(t)$, highlighting feedback from water nanostructures to molecular states**.** Figure 6 shows this scenario.

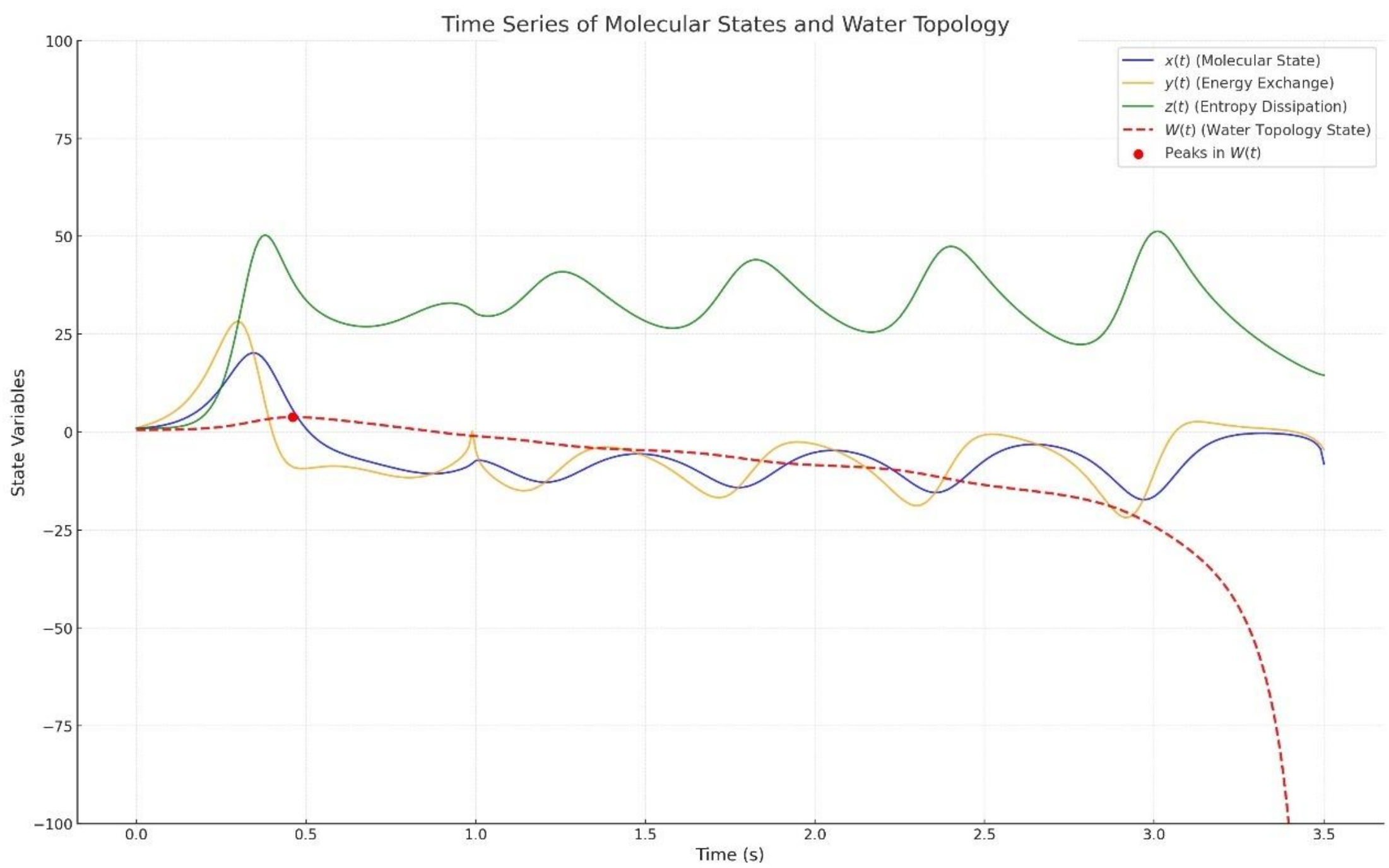

Figure 6

In this Figure is represented the chaotic oscillation in $x(t)$, $y(t)$ and $z(t)$. Molecular states exhibit chaos, with: a) amplitudes modulated by $W(t)$ as peaks and dips in $W(t)$ cause corresponding shifts in $x(t)$, $y(t)$, $z(t)$; b) feedback between states, as the oscillations show complex interdependencies driven by energy input ((t)\) and water topology feedback. Furthermore, Figure 6 shows nonlinear evolution of $W(t)$.

As a matter of fact, $W(t)$ evolves with distinct peaks and dips, acting as a stabilizing force for molecular chaos. Moreover, peaks in $W(t)$ are marked with red dots to emphasize critical moments where water topology strongly interacts with molecular states. To explore bifurcations in the water-modulated chaotic system, we analyze how varying a control parameter (e.g., $\Delta\rho$, the amplitude of water-driven modulation) influences the system's behaviour. Bifurcations occur when a small change in a parameter causes a qualitative shift in the system's dynamics, such as: a) transition from fixed points to periodic behaviour; b) transition from periodic behaviour to chaos.

Therefore, let us perform the following steps: a) fixing key parameters $\rho_o$, $\sigma$ and $\beta$ and other constants; b) varying $\Delta\rho$ (modulation amplitude); c) generating a bifurcation diagram by plotting the long-term behaviour of $x(t)$ (or $z(t)$ as $\Delta\rho$ is varied). Figure 7 shows this scenario, also to identify bifurcations (e.g., transitions to chaos).

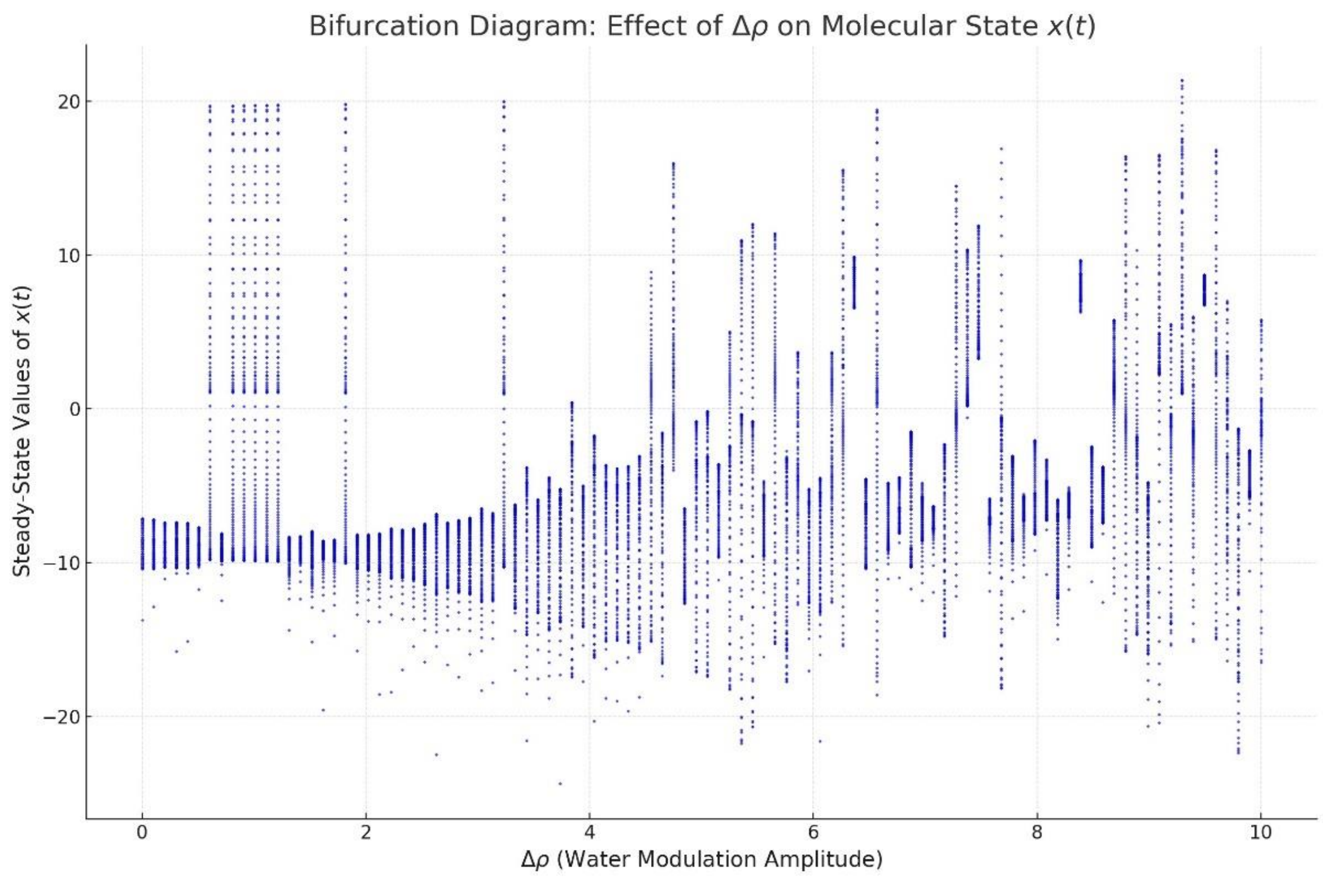

Figure 7

At bifurcation thresholds**,** water nanostructures reorganize dynamically, acting as carriers of information that dictates how molecular states evolve. Information is encoded in water topological fluctuations, which influence system parameters (e.g., energy gradients or entropy dissipation).

Let us examine thresholds by analyzing specific values of $\Delta\rho$ to identify how water- driven information modulates the system state. Figure 8 shows a simulation at key thresholds $\Delta\rho$ =2,5, and 8 (Figure 8).

In this perspective, one can assume that water topology encodes information via the hydrogen bond networks, as water nanostructures encode environmental cues by forming and breaking bonds dynamically, influencing molecular stability. And also, because information flow through water topology modulates entropy dissipation pathways, ensuring stability in biological systems. In the real world, water topology stabilizes intermediate states, encoding folding pathways and ensuring functional conformations. Actually, active site hydration is a dynamic information network, optimizing substrate alignment and transition state stabilization (Ball, 2008; Kovács et al., 2005).

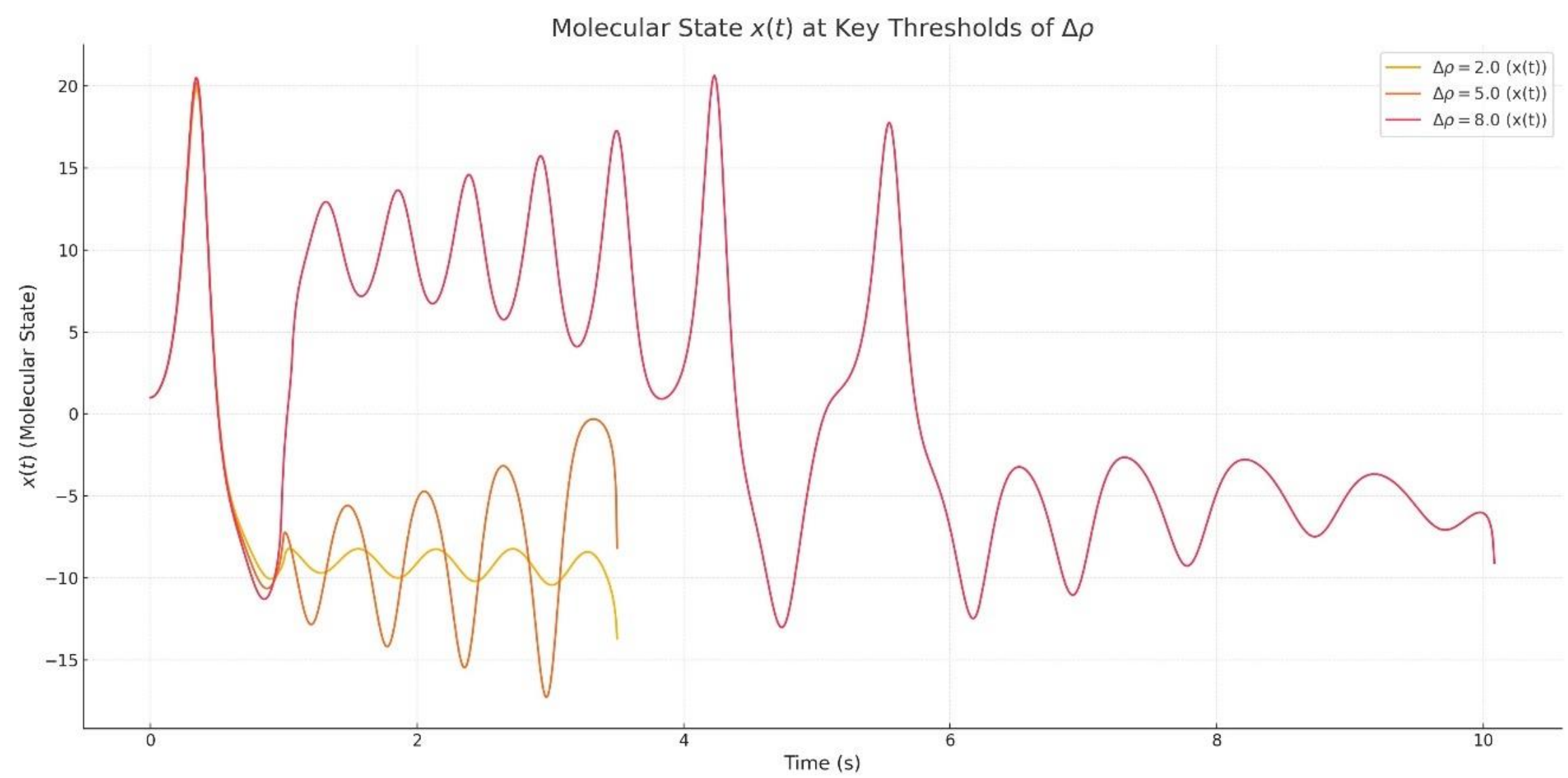

Figure 8

This description, apparently, lacks a “movens” able to explain how life emerged and is organized while is developing. In this perspective, informational Prigogine’s-like dissipative structure may evolve.

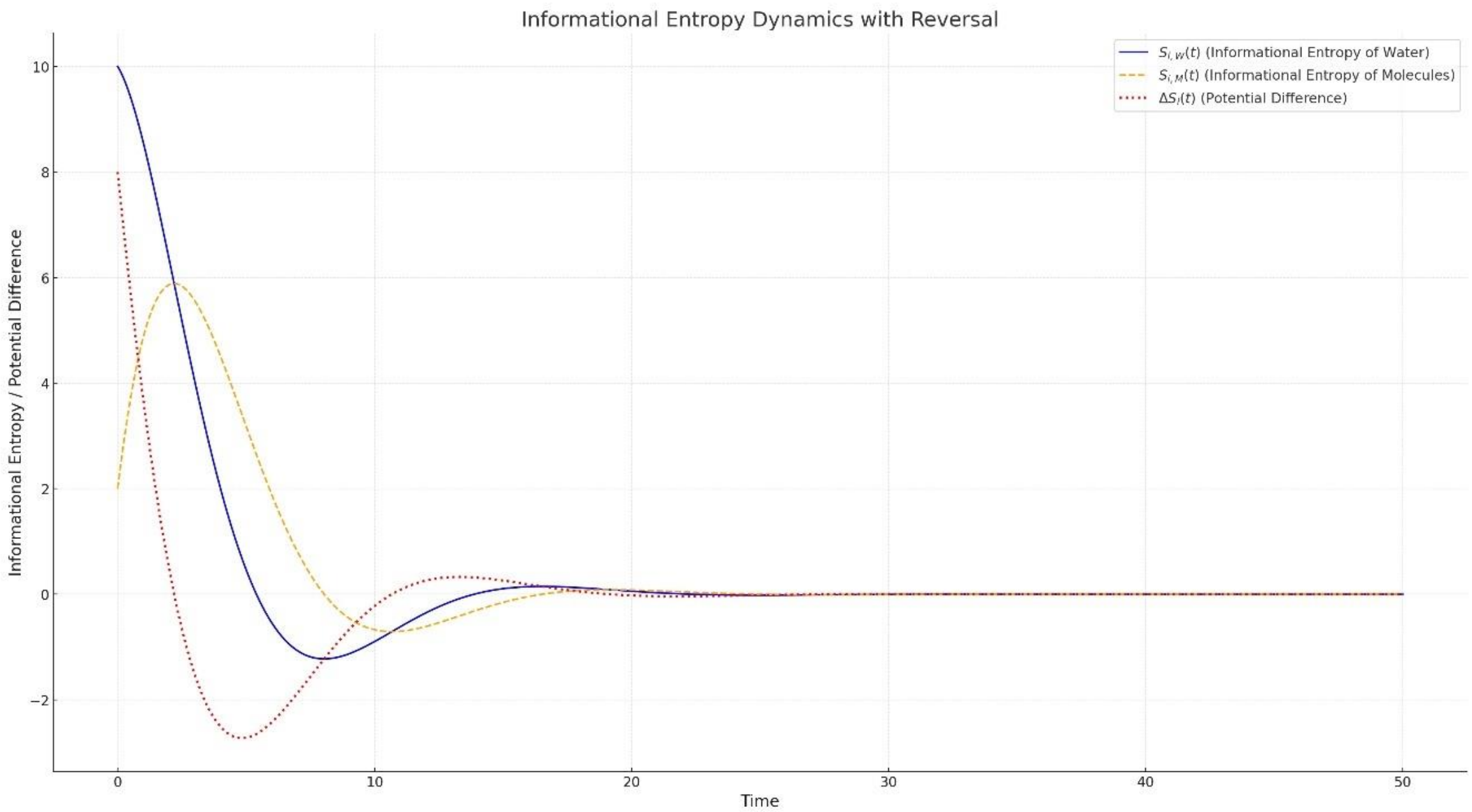

Figure 9

**Prigogine's-like informational dissipative structures move forward life as an event.**

The model forwarded by Gonçalves, shows some bias.

The definition of life as "persistent creation and dissipation of chemical disequilibria" is self-referential and may fail to exclude non-living systems that mimic such behaviours temporarily.
While electron bifurcation is fascinating, its experimental synthesis and relevance to prebiotic chemistry remain speculative. Obviously, the paper dismisses studies supporting genetics-first or membrane-first theories without adequately addressing their strengths or experimental validations. Moreover, by critiquing Darwinian evolution, the paper risks creating a straw-man argument. While Darwinian evolution focuses on populations and heritable traits, the paper appears to conflate it with simple self-replication. We are persuaded that the strong focus on hydrothermal vents and redox gradients biases the discussion, potentially limiting the consideration of alternative environmental conditions for life origin. And finally, the paper assumes that all aspects of life can be reduced to physical and thermodynamic principles, which might overlook emergent biological properties.

A Prigogine's-like informational (Shannon) dissipative structure holds the ability of thermodynamic dissipative systems (with which it is intertwined) to evolve as autopoietic systems, accounting on water and Landauer principles.

One of the major hallmarks of those puzzling living objects called organisms is the ability to reproduce parts of the system and/or the system as a whole, besides to form ordered dynamics, assessing, therefore, typical features belonging to dissipative phenomena (García-Valdecasas, 2022; Mingers, 1989). Actually, the term autopoiesis, derived from the Greek words *auto-* (self) and *poiesis* (creation), describes the process by which a system continuously regenerates and maintains itself. Coined by Humberto Maturana and Francisco Varela in the 1970s, the concept was originally developed to explain the self-organizing and self-sustaining nature of living systems, particularly in biological contexts (Maturana and Varela, 1980; Luisi, 2003; Bertschinger, 2008; Beer, 2015).

In previously published papers, we forwarded the idea that a kind of entropic potential, of informational nature, existing and developing (due to Prigogine's dissipative dynamics) between water topology and the developing chemical mass, i.e., molecules, forced water to regain its high degree of informational freedom (described as degrees of freedom) and as the informational entropy of the event space characterized by molecules and their thermodynamics (defined as the Boltzmann space) increased and was reduced by dissipative dynamics (reiterated, therefore), the difference in informational entropy potential between water (called the Shannon space) and molecules, decreased (Chirumbolo and Vella, 2021; Chirumbolo and Vella, 2023). This phenomenon reversed the transfer process between water and molecules (see later), causing the molecules to break down and bringing the system back to equilibrium. According to this theory, the accumulation of "imperfections," "errors," or "background noise", arising from the informational reiteration linked to Prigogine's dissipative dynamics, within these cyclical and repeated dynamics, increased this entropic potential in favour of the Boltzmann space. This forced water to confine itself, giving the system, now biological, an indefinite duration before the excess of order and reduction of complexity eventually returned the system to thermodynamic equilibrium, that is, the death of Prigogine's dissipative system (Chirumbolo and Vella, 2021; Chirumbolo and Vella, 2023).

The concept of autopoiesis, namely self-organization, self-maintenance, and operational closure, maps intriguingly onto the dynamic and multifaceted behaviour of water topology. Water, as a medium, is fundamental to the emergence and sustenance of autopoietic systems, particularly in biological contexts. Interestingly, we could forward that information arises when a dissipative phenomenon occurs, as information, which is closely linked with the term "memory", might be interpreted as a reiterated event, rather than a fixed property of something. In this sense, autopoiesis might be reported as a way to create and maintain information.

In our previously detailed model, water molecules form transient, dynamic networks via hydrogen bonding. These networks exhibit combinatorial complexity and reorganize continuously. It is arguable that this dynamic reorganization is akin to the self-regenerating processes central to autopoiesis. Actually, the cooperative interactions among water molecules (type II degrees of freedom) provide a cohesive, self-regulating substrate and these degrees of freedom underpin the informational and energetic exchanges needed for the self-maintenance of autopoietic systems (Chirumbolo and Vella, 2021; Chirumbolo and Vella, 2023). Besides the emergence of dissipative structures, in biological systems, the formation of cell membranes relies on the role of water in organizing amphiphilic molecules (e.g., lipids) and it is possible to speculate that water topology determines the stability and flexibility of these membranes, which act as autopoietic boundaries. Moreover, water topology facilitates selective permeability, allow- ing the exchange of nutrients, waste, and signals while maintaining structural and functional integrity. Autopoietic systems manage entropy by converting external energy into organized internal states. Water topology participates by **r**educing thermodynamic entropy inasmuch water structures stabilize molecular interactions, minimizing disorder, and also by increasing informational entropy, as water's combinatorial freedom encodes environmental and systemic informa- tion. A possible idea is therefore that water topology dissipates informational entropy to maintain the operational closure of autopoietic systems, consistent with Landauer's principle.

A possible modelling of the interplay of autopoiesis and water topology is given by:

(13) $$\frac{dW}{dt} = -\lambda W + \kappa(I_{system} - I_{environment})$$

where:

$W$ is the water topology state;

$I_{system}$ is the information entropy of the autopoietic system,

$I_{environment}$ is the external information flux.

In this model water topology modulates energy dissipation and entropy production:

(14) $$\frac{dS_{system}}{dt} = -\alpha W + \beta I_{environment}$$

This model is too simply to shed light on the dynamic of autopoiesis in living matter. More than informational entropy, a first step is to introduce the concept of "degree of freedom", namely the probability of a particle (as a possible informative event) to make relationships with other particles (type I or typeIDFs) or to enhance its intrinsic information if joined or inter-related to other particles (type II or typeIIDFs).

In a first step, water topology, which we can indicate as the Shannon space, i.e., the event space where informational dynamics are of paramount importance to lead forward the life, starts with the highest typeIIDFs and high thermodynamic entropy ($S_{t,W}$), whereas single and separate molecules, moving with gas-like be- haviour (high $S_{t,M}$), (which altogether represent the so-called Boltzmann space) have high typeIDFs but lowest typeIIDFs, therefore water topology transfers informational entropy ($S_{i,W}$) to molecules, reducing the burden of its typeIIDFs.

In the Boltzmann space, we start with lower typeIIDFs (higher typeIDFs), low informational entropy ($S_{i,M}$), and high thermodynamic entropy ($S_{t,M}$). The Boltzmann's space gains typeIIDFs from water, reducing its thermodynamic entropy and creating molecular order. As molecules (Boltzmann space) accumulate typeIIDFs and water topology loses them, the informational entropy gradient ($\Delta S_i = S_{i,W} S_{i,M}$) decreases. Once $S_i$ reverses, the process is compelled to move toward thermodynamic equilibrium, halting Prigogine's dissipative dynamics, as the Boltzmann's space is enabled to transfer typeIIDFs to water, compelling ordered and polymeric molecules to break down.

The informational entropy dynamics for water topology (Shannon's space) ($S_{i,W}$) is characterized by the event that water loses its typeIIDFs to molecules:

$$\frac{dS_{i,W}}{dt} = -\beta S_{i,W} \frac{S_{i,M}}{S_{i,W}} \tag{15}$$

Whereas for molecules (Boltzmann's space) ($S_{i,M}$) the scenario is represented by the gain of typeIIDFs from water:

$$\frac{dS_{i,M}}{dt} = \alpha S_{i,W} \left(1 - \frac{S_{i,M}}{S_{i,W}}\right) \tag{16}$$

As regarding thermodynamic entropy, in the Shannon's space (water topology) thermodynamic entropy increases as water loses informational entropy (typeIIDFs).

$$\frac{dS_{t,W}}{dt} = \gamma \frac{dS_{i,W}}{dt} \tag{17}$$

whereas, thermodynamic entropy in the Boltzmann's state decreases (starting orthodox Prigogine's dissipative dynamics) as molecules gain order (typeIIDFs).

$$\frac{dS_{t,M}}{dt} = -\delta \frac{dS_{i,M}}{dt} \tag{18}$$

The Delta Informational Entropy (DIE) ($\Delta S_i = S_{i,W} S_{i,M}$): a) drives the flow of typeIIDFs; b) reversal occurs when $S_i \rightarrow 0$. Figure 9 describes this modelled scenario. Fundamentally, water topology acts as a chaperone, transferring typeIIDFs to molecules, which reduces molecular thermodynamic entropy and creates order. When the informational entropy gradient ($S_i$) tends to vanish, the system loses its driving force for entropy exchange, compelling it toward thermodynamic equilibrium. The system behaviour models the "death" of the dissipative structure as equilibrium halts entropy flow.

Life emerges when Prigogine's dissipative dynamics force the Boltzmann's space to go ahead in reiterated structures and functional cycles, cumulating therefore underneath imperfections (finalized as chaotic variability) about reiterated structures and/or functions.

This second scenario models the evolution of a complex system where disquisotropic entropy (4,14) ($S^d_{i,M}$) accumulates in the Boltzmann space (molecules) due to the imperfections, cyclicity, and structural reiterations. This underground entropy drives a reversal of entropy flow, compelling water to become confined and highly ordered, pushing the system far from thermodynamic equilibrium.

Therefore, this scenario describes the emergence of life and the progression of the system driven by Prigogine's dissipative dynamics of informational nature. To model this, recently we introduced the concept of disquisotropic entropy as an "underground" accumulation of informational entropy ($S_{i,M}$) in the molecular space (Boltzmann space), resulting in a reversal of the entropy gradient ($S_i$) in favour of molecules.

The mathematical framework considers: a) water topology, i.e., the informational entropy of water ($S_{i,W}$) which loses typeIIDFs during the initial phase but gains them as molecules accumulate underground entropy (disquisotropic entropy) and transfer it back; b) the informational entropy of molecules ($S_{i,M}$), which gains typeIIDFs initially as water acts as a chaperone and accumulates disquisotropic entropy ($S^{disq}_{i,M}$) underground during cycles. In this context, the thermody- namic entropy is inversely proportional to the degree of informational entropy, inasmuch molecules decrease in thermodynamic entropy as they gain order and water increases in thermodynamic entropy during the initial phase, followed by stabilization as confined water emerges. The system turns to transitions to a state far from equilibrium as water becomes confined, sustaining complexity and functional cycles until over-ordering or loss of complexity drives the system back to thermodynamic equilibrium.

Water ($S_{i,W}$) gains typeIIDFs from disquisotropic entropy in molecules:

(19)
$$\frac{dS_{i,W}}{dt} = \beta_1 S_{i,M}\left(1 - \frac{S_{i,W}}{S_{1,M}}\right) + \varphi S_{1,M}^{disq}$$

Molecules entropy ($S_{i,M}$) accumulates underground (disquisotropic) entropy ($S^{disq}_{i,M}$):

(20)
$$\frac{dS_{i,M}}{dt} = \alpha_1 S_{1,W}\left(1 - \frac{S_{i,M}}{S_{i,W}}\right) - \varphi S_{1,M}^{disq}$$

whereas disquisotropic entropy ($S^{disq}_{i,M}$) accumulates underground in molecules during cyclic reiterations:

(21)
$$\frac{dS_{i,M}^{disq}}{dt} = \sigma S_{i,M}\left(1 - \frac{S_{i,M}^{disq}}{S_{i,M}}\right)$$

In this context the thermodynamic entropy should behave as follows:

a) For water, stabilizes as water becomes confined:

(22) $$\frac{dS_{t,W}}{dt} = \gamma \frac{dS_{i,W}}{dt} - \lambda S_{i,W}$$

For molecules, thermodynamic entropy increases as molecules over-order or lose complexity

(23) $$\frac{dS_{t,M}}{dt} = \delta \frac{dS_{i,M}}{dt} + \omega S_{i,M}^{disq}$$

In this context, the delta entropy (DIE), i.e., $\Delta S_i = (S_{i,W} - S_{i,M})$, initially favours water but transitions to favour molecules as $S^{disq}_{i,M}$ grows.

This ongoing cumulation of imperfections respect the rigid cyclic similarity, or "errors", creates chaotic regimes. To address this point, it should be measured key metrics such as the rate of disquisotropic entropy accumulation and the energy dissipation governed by Landauer's principle, and moreover it should be assessed how these factors influence system dynamics and lifetime. Therefore, I simulated longer-term behaviour, including possible perturbations (e.g., external energy inputs or molecular instability). And I explored cases where confined water undergoes structural disruptions or over-ordering. The involvement of chaos could be highlighted by investigating how chaotic variability in entropy flows ($S_{i,W}$, $S_{i,M}$, and $S^{disq}_{i,M}$ ) impacts the stabilization and collapse of the system. And furthermore, I introduced chaotic or stochastic perturbations to entropy dynamics and assessed their role in sustaining or destabilizing complexity.

For the numerical quantification, it was introduced the rate of disquisotropic entropy growth $\frac{dS_{i,M}^{disq}}{dt}$

a) it quantifies how imperfections and variability accu- mulate in the molecular space, and b) indicates the rate at which typeIIDFs are added to molecules.

Moreover, to calculate the energy dissipation via the Landauer's principle, it should be computed the thermodynamic entropy produced during entropy transfer:

(24) $$\Delta S_t = k_B \ln \Delta I$$

where $\Delta I$ is the informational entropy erased.

In this scenario it should be assessed how long the system remains far from equilibrium before returning to thermodynamic equilibrium. Chaos influences the system by a) adding chaotic or stochastic components to entropy dynamics (e.g., sin $k \cdot t$, random noise). Exploring how small deviations in entropy flows amplify over time. Furthermore, chaos disrupts regular cycles, extending system persistence far from equilibrium. And finally, chaotic feedback prevents stagnation, allowing the system to explore novel configurations.

Let us introduce chaotic terms into the equations

(25) $$\frac{dS_{i,W}}{dt} = \beta_1 S_{i,M}\left(1 - \frac{S_{i,W}}{S_{1,M}}\right) + \varphi S_{1,M}^{disq} + \kappa \sin(k \cdot t)$$ instead of Eq. (19)

(26) $$\frac{dS_{i,M}^{disq}}{dt} = \sigma S_{i,M}\left(1 - \frac{S_{i,M}^{disq}}{S_{i,M}}\right) + \eta(t)$$ instead of Eq. (21)

where η(t) is a random noise term. Figure 10 shows finally all this described issue, whereas Figure 11 shows the growth rate of disquisotropic entropy.

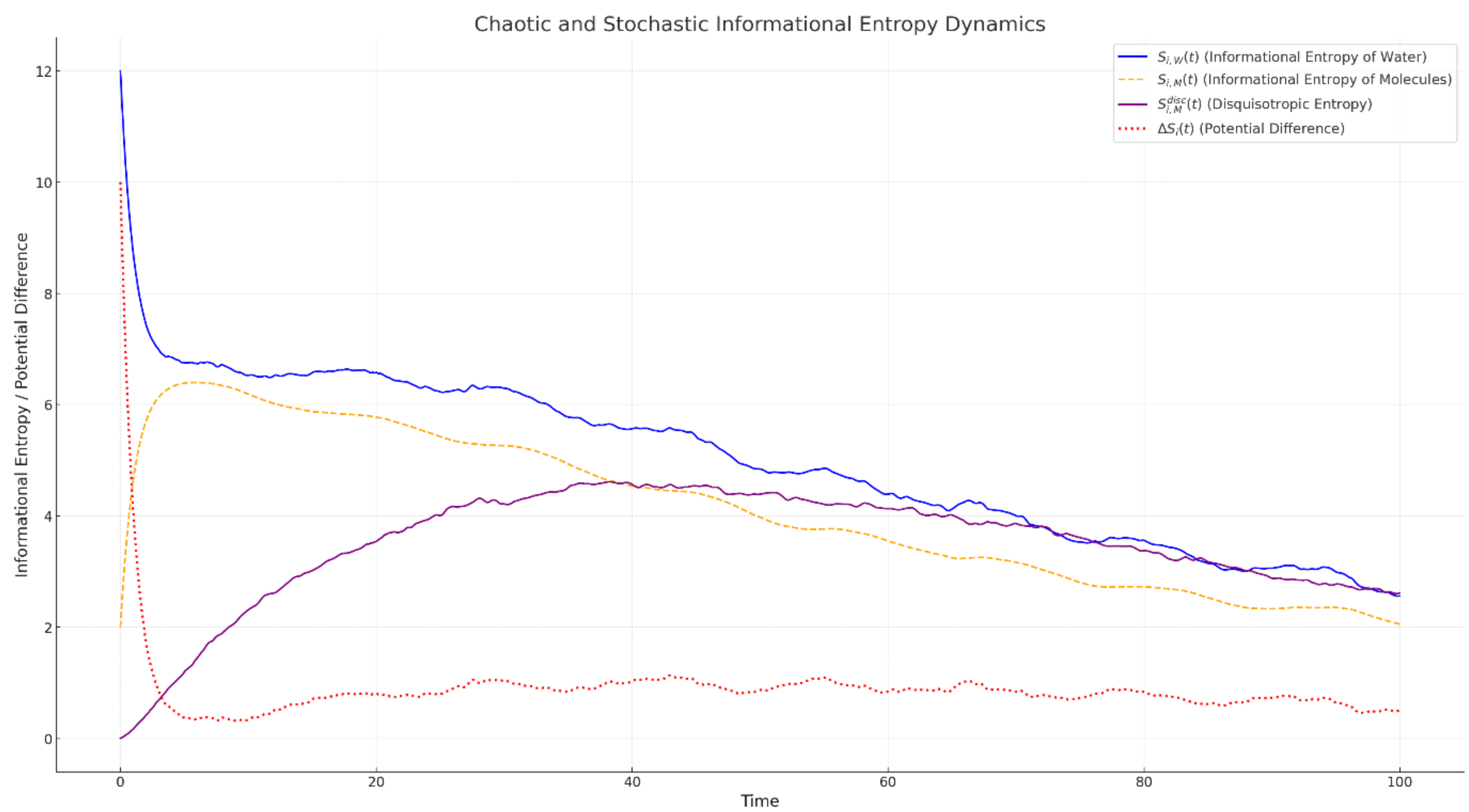

Figure 10

The noise is both causative of and produced by disquisotropic entropy. Noticeably, Figure 11 shows that this kind of hidden informational entropy stabilizes the system far from the thermodynamic equilibrium. To refine the model, we could allow the noise amplitude () to vary based on the state of the system to simulate environmental changes. And also introduce adaptive constraints to entropy gradients, dynamically modulating chaos intensity without violating thermodynamic principles. Therefore, we should quantify the sensitivity of the system to initial conditions by calculating the Lyapunov exponent, which measures chaotic behaviour. And moreover, we can visualize entropy dynamics in a multi-dimensional phase space to better understand attractors and chaotic trajectories. The model can be improved by introducing periodic or random external perturbations to entropy flows and study their impact on system cycles. Furthermore, we may explore how changes in initial energy or entropy reservoirs impact the system's longevity and also, we may simulate different chaotic intensities (e.g., varying, κ,σ or k) to assess their influence on system stability.

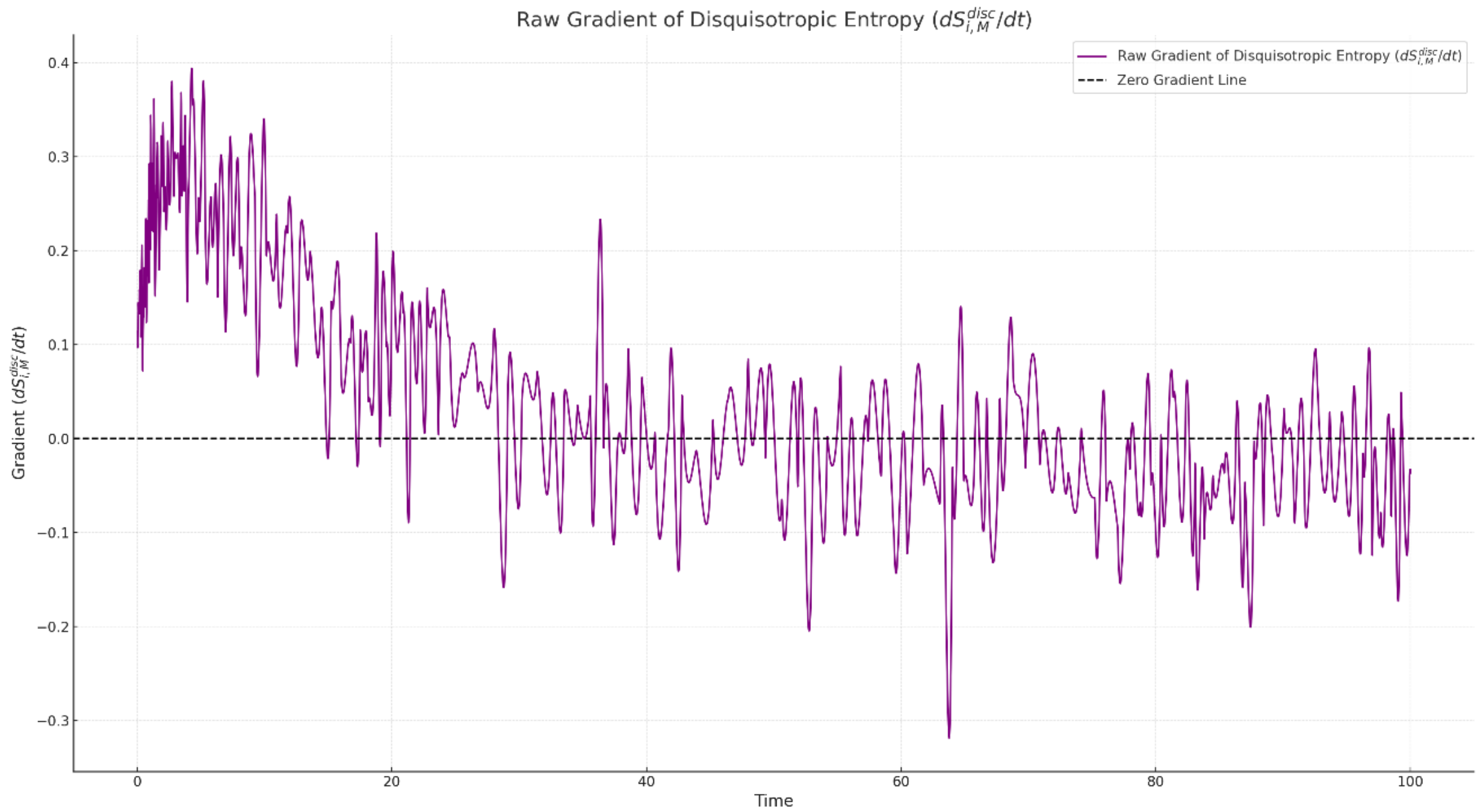

Figure 11

Figure 12 is a result of this evaluation.

Chaos enriches system dynamics, maintaining entropy flows and sustaining cycles. The sensitive dependence on initial conditions ensures variability, delaying equilibrium. Furthermore, adaptive noise and constraints align the model more closely with physical phenomena and finally chaotic oscillations are modulated by disquisotropic entropy growth and external influences.

Through refined models incorporating chaotic and stochastic influences, it has been shown the intricate feedback mechanisms that sustain cycles of entropy exchange. Using a novel framework that integrates Landauer's principle, it could be quantified the thermodynamic cost of entropy flows, elucidating the energy dissipation limits that govern the system's lifespan. Chaotic trajectories in phase space and positive Lyapunov exponents confirm the sensitive dependence on initial conditions, while refined energy dissipation models ensure alignment with thermodynamic principles.

The findings extend Prigogine's theory of dissipative structures by introducing disquisotropic entropy as a key factor in prolonging system persistence, offering a plausible explanation for the transition from physical to biological regimes.

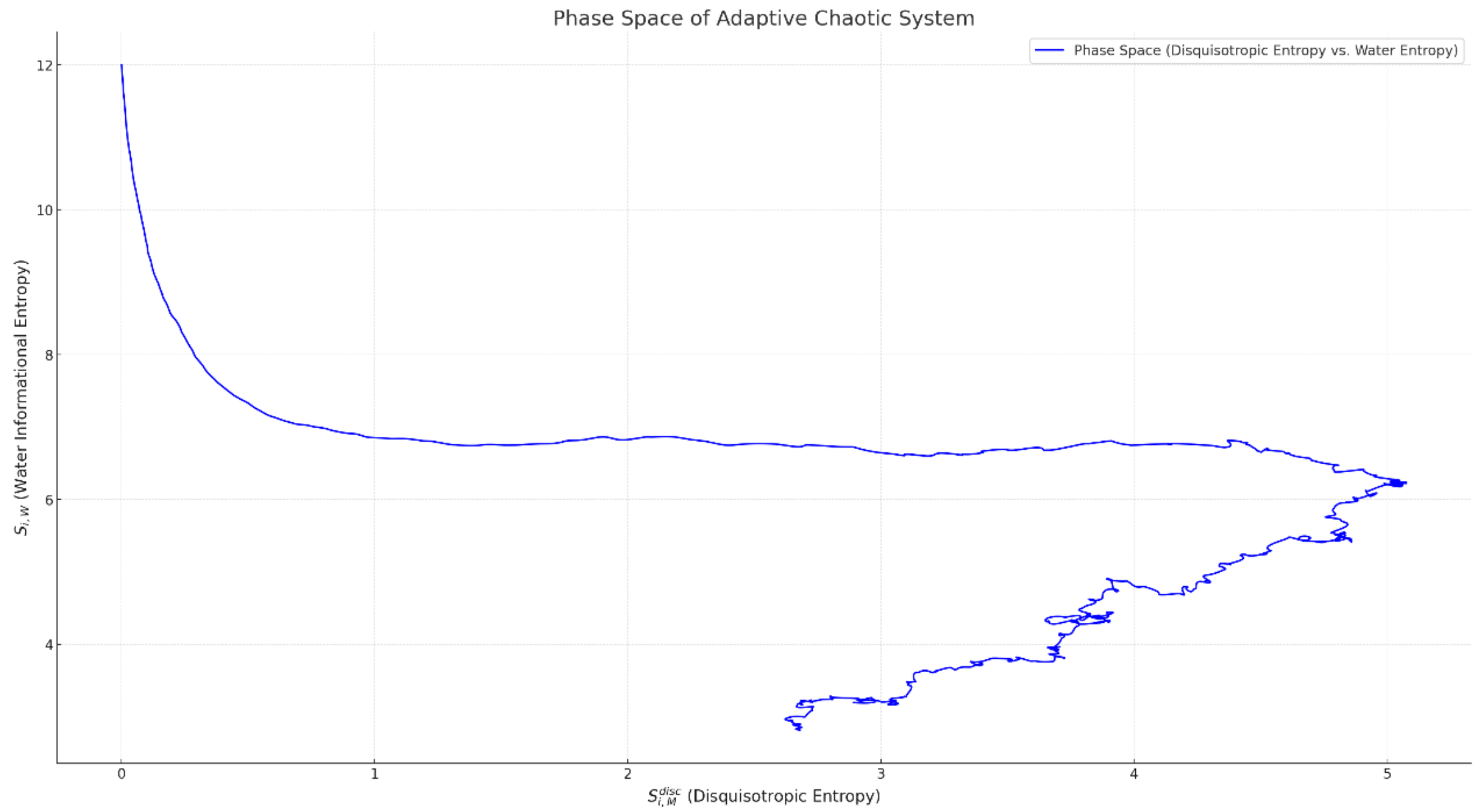

Figure 12

## Conclusions

The model proposed by Gonçalves, while innovative, may confuse readers by implying that systems like chemical oscillators (e.g., Belousov-Zhabotinsky reactions) qualify as living, without a reappraisal of the concept of information. Furthermore, the claim that the proposed principles universally apply to all life forms might be premature without empirical validation or application to extra-terrestrial contexts. Yet, the model we described here appears quite generalist, in its fundamental ground.

This exploration of chaotic dynamics, entropy flows, and Prigogine's dissipative systems illuminates critical mechanisms underlying the stabilization, evolution, and eventual collapse of far-from-equilibrium systems. By integrating theoretical constructs with refined models, we have uncovered insights that connect chaos, disquisotropic entropy, and Landauer's principle to the emergence and persistence of complexity in physical and potentially biological systems. There are some key findings to take and bring home. The interplay between water topology (Shannon space) and molecular systems (Boltzmann space) creates a dynamic cycle where water acts as a kind of informational chaperone, guiding molecular order. The accumulation of imperfections and variability within molecular systems acts as a hidden dynamic reservoir of informational entropy, stabilizing the system far from thermodynamic equilibrium and extending its lifespan. Chaos enriches the system's adaptability, delaying equilibrium by creating irregular cycles and feedback loops. The positive Lyapunov exponent confirms chaotic behaviour, highlighting its role in sustaining cycles through continuous reorganization of entropy flows. Moreover, chaotic trajectories in phase space reveal the intricate interplay of disquisotropic entropy, water topology, and thermodynamic constraints. Again, energy dissipation, constrained by Landauer's principle, governs the limits of system longevity. And finally, the observed zero-crossings in dissipation highlight the sensitivity of chaotic systems to stochastic influences and modelling nuances, emphasizing the importance of refining energy dissipation models.

As entropy gradients diminish and energy reservoirs deplete, the system inevitably returns to thermodynamic equilibrium, halting the cycles that sustain complexity.

This study demonstrates that chaos is not inherently destabilizing; rather, it enables resilience and adaptability, essential features for systems seeking to maintain order in fluctuating environments. The identified feedback mechanisms, cyclic entropy flows, and interplay between order and chaos provide a plausible framework for understanding how molecular systems might transition from phys- ical to biological regimes. And finally, the introduction of disquisotropic entropy as a stabilizing factor highlights a novel pathway for extending the lifespan of dissipative systems, offering insights into the maintenance of complexity in both natural and artificial systems.

These principles not only shed light on the origins of complexity in living systems but also have applications in synthetic biology, energy management, and the design of adaptive systems. This work establishes a foundation for exploring how chaos and entropy can be harnessed to sustain complexity, advancing our understanding of life-like processes in both natural and engineered systems.

## Figure legends

Figure 1. 3D plots depicting the behaviour of the Lorenz system at different Rayleigh numbers (ρ\rho): A) ρ (rho) = 10.0 (Low Energy Input): The systemconverges to a stable fixed point, representing a non-chaotic dissipative structure; B) ρ (rho) = 24.74 (Bifurcation Point): the system exhibits periodic or quasi- periodic behaviour, indicating the onset of instability but not yet full chaos. C) ρ(rho) = 28.0 (High Energy Input): the system becomes chaotic, with a highly sensitive trajectory that never repeats, exemplifying chaotic dynamics. This progression illustrates how a dissipative structure becomes chaotic as energy input (ρ\rho) increases. Plotted with Python code (Python 3.9) in Matlab (R2022a) environment.

Figure 2. 3D Trajectory Plot. The modified Lorenz system shows highly complex and dynamic behaviour. The trajectory no longer exhibits the typical Lorenz attractor structure, as ρ(t) varies with time and the system state, creating non-stationary chaos. Plotted with Python code (Python 3.9) in Matlab (R2022a) environment.

Figure 3. Time Series of ρ(t) and z(t). Here, ρ(t)\ (blue curve) oscillates dynamically due to water-mediated effects. The system state z(t) (orange curve) evolves chaotically, influenced by ρ(t). The interplay between dynamic energy input ρ(t) and system response z(t) reflects how water molecules condition the energy flow in the system. Plotted with Python code (Python 3.9) in Matlab (R2022a) environment.

Figure 4. 3D Trajectory with Water Topology Effects. The trajectory displays enhanced chaotic behaviour compared to the original Lorenz attractor. The addition of dynamic ρ(t) and stochastic noise from water topology causes the sys-

tem to explore a broader phase space, with irregular oscillations and transitions. Plotted with Python code (Python 3.9) in Matlab (R2022a) environment.

Figure 5. Time Series of State Variables (x,y,z) and Dynamic ρ(t). The state variables (x(t),y(t), z(t)) exhibit non-stationary, chaotic oscillations, influenced by the dynamic and fluctuating (t). The red dashed curve (ρ(t)) shows how water-mediated effects modulate the energy input dynamically, coupling with the system's evolution. Plotted with Python code (Python 3.9) in Matlab (R2022a) environment.

Figure 6. The Figure shows molecular states (*x(t)*, *y(t)* and *z(t)*), which represent molecular dynamics, energy exchange, and entropy dissipation, respectively. Chaotic oscillations are clearly visible, with amplitudes modulated by W(t). The water topology state (W(t)) evolves nonlinearly and is plotted with a dashed red line for emphasis. Peaks in W(t) are marked with red dots, highlighting moments of pronounced feedback between water nanostructures and molecular states. Distinct colors and line styles improve readability. The y-axis scaling has been adjusted to ensure all variables are visible, avoiding compression near zero. Actually, the y-axis range is expanded to 100-100, offering a comprehensive view of potential outliers or extreme values. This visualization emphasizes the full scope of dynamic variability within the system. Plotted with Python code (Python 3.9) in Matlab (R2022a) environment.

Figure 7. The bifurcation diagram illustrates how the steady-state behaviour of the molecular state *x(t)* evolves as Δρ (the amplitude of water modulation) increases. For low Δρ, the system exhibits single attractor behaviour (fixed points or simple periodic oscillations). For intermediate Δρ bifurcations emerge, indicating the system transitions to periodic oscillations or quasi-periodic behavior. Water modulation introduces oscillatory feedback, altering molecular dynamics. For high Δρ the system becomes chaotic, with many steady-state values for *x(t)*, highlighting sensitive dependence on initial conditions. This demonstrates how strong water-driven modulation destabilizes the system. Key Insights: a) the diagram highlights critical thresholds (Δρ) where the system transitions from order to chaos. Water topology, through Δρ, modulates the molecular state behaviour, driving complex dynamics and emergent chaos. Plotted with Python code (Python 3.9) in Matlab (R2022a) environment.

Figure 8. Entropy dissipation at key thresholds. At low Δρ=2 *z(t)* shows smooth oscillations, indicating stable energy dissipation. Information flow through water topology maintains order in the system, enabling predictable entropy dissipation. At Intermediate Δρ=5**, t**he oscillations in *z(t)* become more irregular. Water-driven information introduces variability, suggesting adaptive dissipation in response to environmental fluctuations. At high ΔΔ=8**,** *z(t)* displays chaotic fluctuations, with no clear periodicity. Water topology acts as a chaotic information generator, enhancing system complexity and dissipative behaviour. Plotted with Python code (Python 3.9) in Matlab (R2022a) environment.

Figure 9. Plot of the informational entropy dynamics. For water $S_{i,W}(t)$ is initially high due to abundant typeIIDFs in water topology. Then gradually decreases as water transfers informational entropy to molecules. For molecules, $S_{i,M}(t)$ starts low, reflecting fewer typeIIDFs in molecules. Then, increases over time as molecules gain order from water's typeIIDFs. Therefore, DIE (

($\Delta S_i(t)=S_{i,W}S_{i,M}$**)** decreases steadily, reflecting the diminishing entropy gradient between water and molecules. Then it approaches zero, signaling the reversal point where molecules contain more typeIIDFs than water, driving the system towards thermodynamic equilibrium. Plotted with Python code (Python 3.9) in Matlab (R2022a) environment.

Figure 10. Plot showing the modelling of Prigogine-like informational dissipative system once disquisotropic entropy forces the system to go ahead towards order and complexity far from the thermodynamic equilibrium. It shows the chaotic informational entropy dynamics. Water $S_{i,W}$ displays irregular oscillations due to chaotic and stochastic influences. It gains and loses typeIIDFs dynamically, reflecting the interplay with molecular disquisotropic entropy. On the other hand, molecules $S_{i,M}$ shows fluctuating growth patterns as chaos disrupts regular entropy flow. And disquisotropic entropy ($S^{disq}{}_{i,M}$) grows unpredictably but supports molecular complexity. The potential difference ($\Delta S_i$) shows that irregular fluctuations indicate chaotic exchanges of entropy between water and molecules, extending system persistence. Plotted with Python code (Python 3.9) in Matlab (R2022a) environment.

Figure 11. This plot highlights the chaotic accumulation of molecular imperfections and variability.

The rate fluctuates dynamically, emphasizing how chaos delays equilibrium and sustains complexity.

Note that the fluctuations are around zero (straight line), showing a stable dynamic of the system. Plotted with Python code (Python 3.9) in Matlab (R2022a) environment.

Figure 12. The computed Lyapunov exponent is approximately 0.12. A positive Lyapunov exponent confirms that the system exhibits chaotic dynamics, indicating sensitivity to initial conditions and the presence of exponential divergence in trajectories. Furthermore, the phase space plot shows a complex, non-repeating trajectory, consistent with chaotic behaviour. This visualization highlights the intricate interaction between disquisotropic entropy ($S^{disqi,M}$) and water entropy ($S_{i,W}$). Plotted with Python code (Python 3.9) in Matlab (R2022a) environment.